# Machine Learning Enabled Prediction of Cathode Materials for Zn ion Batteries

Linming Zhou[1,2], Archie Mingze Yao[3], Yongjun Wu[1, 4#], Ziyi Hu[1], Yuhui Huang[1, #], Zijian Hong[1,2#]

[1] Lab of dielectric Materials, School of Materials Science and Engineering, Zhejiang University, Hangzhou, Zhejiang 310027, China

[2] Hangzhou Global Scientific and Technological Innovation Center, Zhejiang University, Hangzhou, Zhejiang 311200, China

[3] Department of Mechanical Engineering, Carnegie Mellon University, Pittsburgh, PA, 15213, USA

[4] State Key Laboratory of Silicon Materials, Cyrus Tang Center for Sensor Materials and Applications, School of Materials Science and Engineering, Zhejiang University, Hangzhou 310027, China

Abstract

Rechargeable Zn batteries with aqueous electrolytes have been considered as promising alternative energy storage technology, with various advantages such as low cost, high volumetric capacity, environmentally friendly, and high safety. However, a lack of reliable cathode materials has largely pledged their applications. Herein, we developed a machine learning (ML) based approach to predict cathodes with high capacity (>150 mAh/g) and high voltage (>0.5V). We screened over ~130,000 inorganic materials from the Materials Project database and applied the crystal graph convolutional neural network (CGCNN) based ML approach with data from the AFLOW database. The combination of these two could not only screen cathode materials that match well with the experimental data but also predict new promising candidates for further experimental validations. We hope this study could spur further interests in ML-based advanced theoretical tools for battery materials discovery.

# Corresponding authors

Y. W. (yongjunwu@zju.edu.cn) Y. H. (Huangyuhui@zju.edu.cn)

Z. H. (hongzijian100@zju.edu.cn)

Rechargeable Zn-ion batteries with aqueous electrolytes have been considered as promising alternatives and supplements for the commercialized rechargeable Li-ion batteries, with many advantages such as low cost, high volumetric energy density, environmentally benign, and high safety. [1-6] They are promising for device applications in large-scale grid energy storage and renewable green energy electrochemical storage systems, which are sensitive to cost, space and environmental effects. However, one of the key challenges facing the Zn-ion battery systems is the lack of proper cathode materials with high structure stability, high capacity, and high voltage [7]. To date, manganese oxides [8-11], vanadium oxides [12-17], Prussian blue analogs [18-20], $Mo_6S_8$-based Chevrel phase compounds [21], and some other organic materials [22, 23], etc. have been explored as cathode materials for Zn batteries. Meanwhile, they still suffer from either low capacity, low voltage, or low cycle life, and so on, which severely hindered their practical applications. Discovering and designing reliable new cathode materials are essential for the future development, application, and commercialization of the Zn battery systems.

Recently, the data-driven machine learning (ML) approach has largely renovated the paradigm for materials design and discovery, which opened a new era for the computational materials science community [24, 25]. The combination of large data sets in the materials databases, data-mining technique, structure featurization, and ML approach could not only expand the materials screening scope but also gain unique insights into the complex structure-property and composition-property relationship, using the wealth of the existing computational and experimental materials data [26, 27]. It can achieve a desirable prediction accuracy for materials properties that is comparable to the conventional computational materials tools such as density functional theory (DFT) with a lower computational cost or larger computational length-scale /timescale. One particular example is the recent development of the crystal graph convolutional neural network (CGCNN) [28], a ML tool that featurize the crystal structure as a “graph”, has enabled the accelerated design of crystalline materials with desired properties, while also providing fundamental insights on the elementary/structural contributions to the energetics and properties of the materials. It has been successfully applied to the design of various crystalline materials for specific properties and applications, including the screening of solid electrolytes for dendrite suppression in lithium battery [29], prediction of thermodynamic, mechanical, and electronic properties for 2-D materials [30], the formation energy of compounds [31], and the methane adsorption properties of metal-organic frameworks [32], etc.

Herein, we developed a CGCNN-based ML tool to predict high voltage cathode materials for Zn-ion batteries. Using materials data from two representative materials databases, namely the materials project (MP [33]) and AFLOW [34, 35], we screened over 130,000 inorganic materials and extracted ~60 candidates that are predicted to have high voltage (over 0.5 V vs $Zn/Zn^{2+}$), low toxicity, high abundant and high capacity (over 100 mAh/g). It is discovered that the mixing prediction method which combined both the MP-derived and ML-predicted voltages agrees remarkably well with the experimental measurements. While it also predicted tens of candidates with both high voltage and high capacity that have not been explored yet, which is tabulated for future experimental validations. We hope this study could spur further interests in the experimental discovery of cathode materials for Zn batteries and theoretical screening of other battery materials with highly efficient ML-based tools.

From classic electrochemical theory, the electrochemical reaction for a standard Zn full cell with Zn metal anode can be written as:

$$a\mathrm{Zn}+b\mathrm{X}=\mathrm{Zn}_a\mathrm{X}_b \tag{1}$$

Where $X//Zn_aX_b$ is the cathode material pair before and after intercalation of Zn. The electrochemical potential for the *Zn//X* full cell can be derived from the Gibbs free energy of the reaction (1) $\Delta G_1$, namely:

$$\Delta G_1 = G_{Zn_aX_b} - (aG_{Zn} + bG_X) \tag{2}$$

$$\varphi = \Delta G_1/(2aF) \quad (3)$$

Where $a$, $b$ are the balance factors, $G_{Zn_aX_b}$, $G_{Zn}$ and $G_X$ are the Gibbs free energy of $Zn_aX_b$, Zn and $X$, respectively. $\varphi$, 2, and F are the equilibrium electrochemical potential of the cathode pair, number of electrons involved per Zn atom for the electrochemical reaction (1), and Faraday constants, respectively. Combining Eq. (2) and Eq. (3), it is indicated that the accurate prediction of electrochemical potential $\varphi$ requires reliable calculations of the Gibbs free energy for each material. Traditionally, the free energy of materials in the materials database are obtained from high-throughput calculations, which could give rise to a systematic error that led to large deviations for potential predictions. To overcome this problem, we employ a combined approach with the CGCNN-based ML tool and multiple databases to improve the prediction accuracy while also provide uncertainty qualifications for the predictions.

The schematics of the battery cathode prediction protocol used in this study are given in **Fig.1**. Using the Pymatgen API, we search through the existing ~130,000 inorganic materials in the MP database and obtain ~19,000 pairs of electrodes with Zn-containing compounds ($Zn_aX_b$) and the corresponding material before intercalation of Zn (*X*). Among these, compounds with toxic elements (e.g., Pb, Cd, Tl, Cr, Hg) and radioactive elements (U, Fr, Ra, etc.) are excluded which reduces the candidates to ~9,000 pairs. The electrochemical potential of these cathode pairs with respect to $Zn//Zn^{2+}$ can be obtained by combining equations (2) and (3), with the Gibbs free energy data extracted from the MP database (denoted as $\varphi_{MP}$). Meanwhile, subsequently, a DFT-Solver based on the CGCNN algorithm was trained to predict the Gibbs free energy for all the ~130,000 inorganic materials pool. Then, a total of ~60,000 data instances are taken from the AFLOW database, where ~42,000 data instances are used as training sets (70%), ~9,000 data are chosen as the validation set (15%) and ~9,000 data are employed as testing set (15%). Similarly, the electrochemical potential for the ~9,000 cathode pairs can also be calculated with the Gibbs free energy predicted from the CGCNN-based ML approach, which is denoted as $\varphi_{ML}$. A step further, after filtering out all the positive potential value for the two electrochemical potentials ($\varphi_{MP}$ and $\varphi_{ML}$), we define a mixing prediction potential by calculating the square root of the two predicted potentials, i.e.,

$$\varphi_{mix} = \sqrt{\varphi_{MP}\varphi_{ML}} \tag{4}$$

Further filtering is applied to screen all the candidates that have both high mixing prediction potential (>0.5V) and high capacity (>100 mAh/g). The electrode candidates can be obtained after combining all the compounds that have the same elements and similar *a* to *b* ratio for further experimental materials synthesis and test to validate the prediction.

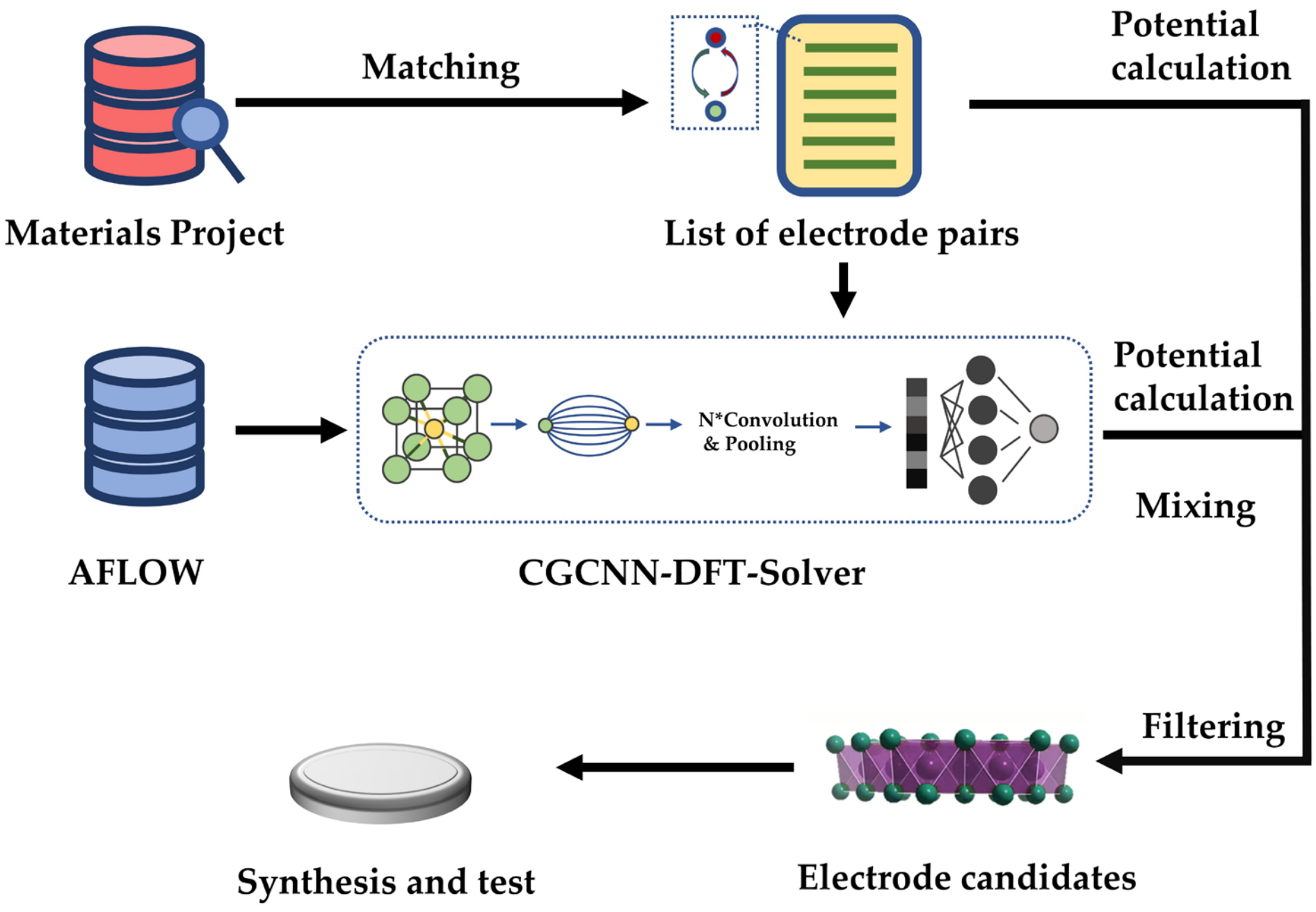

**Fig. 1|** Schematics of the ML-assisted battery electrode material screening protocol. The electrochemical potentials are calculated from Materials Project database and predicted by ML with data from AFLOW materials database. Then, the electrode candidates can be obtained by selecting the high voltage (>0.5 V) and high capacity (>100 mAh/g) compounds.

To validate the CGCNN-based ML DFT-solver, the parity plot for the ML-predicted energies and high throughput DFT calculated energies taken from the two materials databases is given in **Fig.2**. It can be observed that the CGCNN-DFT-solver with training data taken from both MP and AFLOW shows good linearity with the energy data extracted directly from the two materials databases. The coefficients of determination ($R^2$) for both cases are close to 1, indicating very small standard deviations, while the slopes of the lines for both MP and AFLOW are 0.98 and 0.99, respectively, showing excellent correlations between the ML-predicted energies and the energies extracted directly from the databases. The Mean Absolute Error (MAE) is plotted against the machine learning epoch for the training and validation set, showing the saturation of MAE after 100 epochs for both databases, giving rise to a MAE value of ~0.1 eV/atom, consistent with the previous report [28]. This MAE value also indicates that the standard error for the CGCNN-ML model is within the DFT model uncertainty.

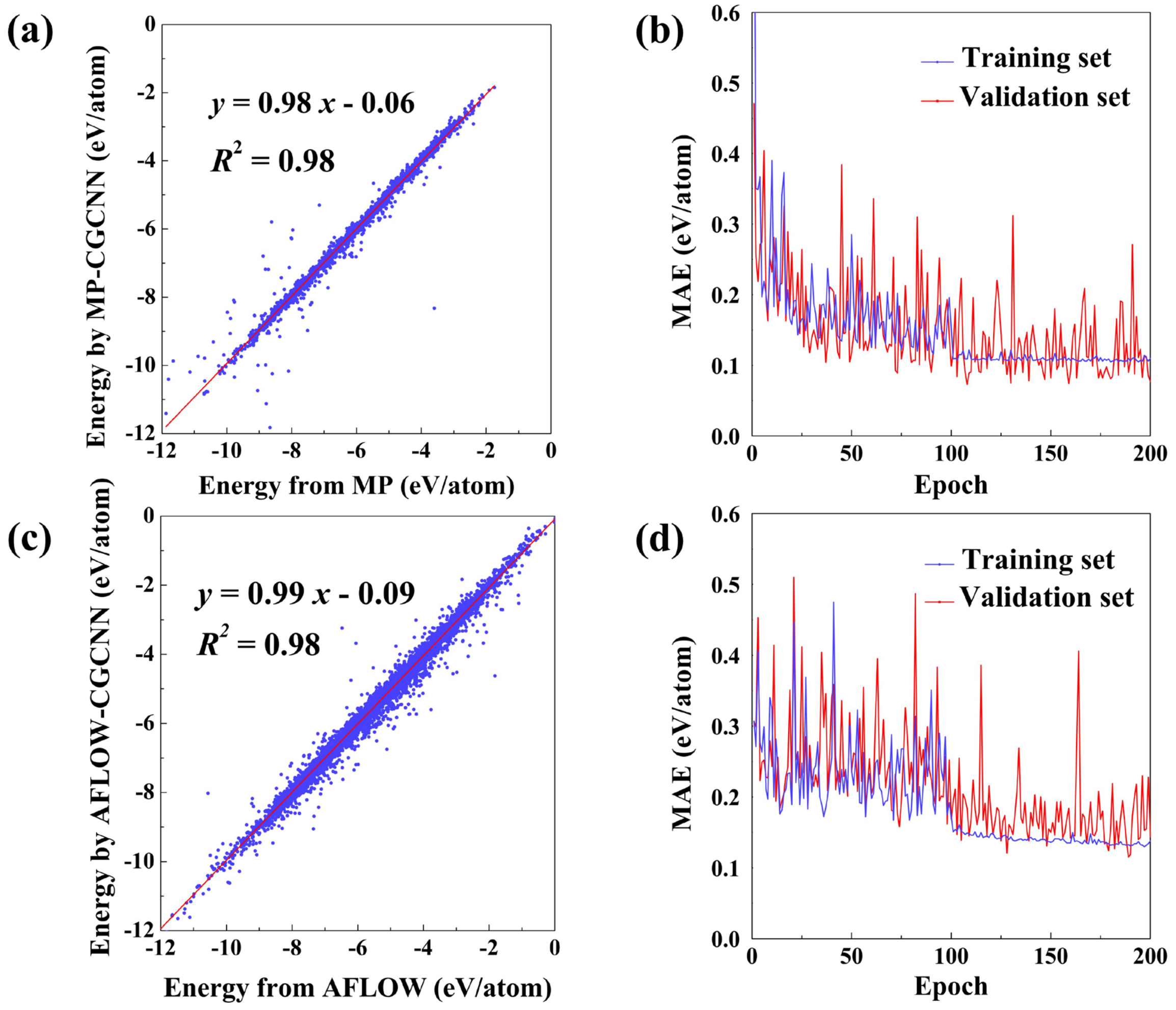

**Fig. 2|** Performance of machine-learned models for the Gibbs free energy. (a) Parity plot for Gibbs free energy predictions with CGCNN using training data from Materials Project. (b) Mean absolute error (MAE) versus training epochs for training set and validation set. (c) Parity plot for Gibbs free energy predictions with CGCNN using training data from AFLOW. (d) MAE vs. training epochs with training data taken from AFLOW database.

The statistics for this study are further given in **Fig.3**. It can be seen that a total of ~9,000 electrode pairs has been screened from the MP database out of ~130,000 inorganic compounds (**Fig.3a**). Around 3/4 of these electrode pairs have a positive electrochemical potential. Notably, the number of candidates decays exponentially with the increasing of electrochemical potential, the majority of candidates show a calculated electrochemical equilibrium potential between 0 and 2 V. A similar trend is observed from the ML-learned potentials with training data from AFLOW (e.g., $\varphi_{ML}$), where most of the electrode pairs have a relatively low voltage between 0 to 1.5V. The low predicted voltage can be understood since the electrochemical potential for $Zn//Zn^{2+}$ (-0.76 V vs. S.H.E.) is much higher than $Li//Li^{+}$ (-3.06 V vs. S.H.E.). Meanwhile, it is shown that the actual statistical distributions for the two cases are not identical. To quantify the agreements between the two potentials, the correlation between $\varphi_{MP}$ and $\varphi_{ML}$ is further plotted in **Fig. 3(b)**. It can be observed that these two potentials show a wide dispersion and week and positive correlation with R-squared of ~0.4 and $\rho_{xy}$ of 0.6, while the slop is 0.41. The weak correlation can be understood since the materials data in the two databases are obtained with different high throughput calculation methods. Meanwhile, the fundamental elementary and structural contributions to the energies should be similar throughout the databases. To further reduce the systematic error from the two calculation methods, the mixing potential $\varphi_{mix}$ is calculated by taking the mean square of $\varphi_{MP}$ and $\varphi_{ML}$, as indicated in Eq.4. The distribution of $\varphi_{mix}$ is demonstrated in **Fig. 3(c)**, which gives rise to almost equal number of pairs in between 0 V and 1.5 V while decays exponentially above 1.5 V. The total number of pairs that exhibit a potential above 0.5 V reduces to ~3,000.

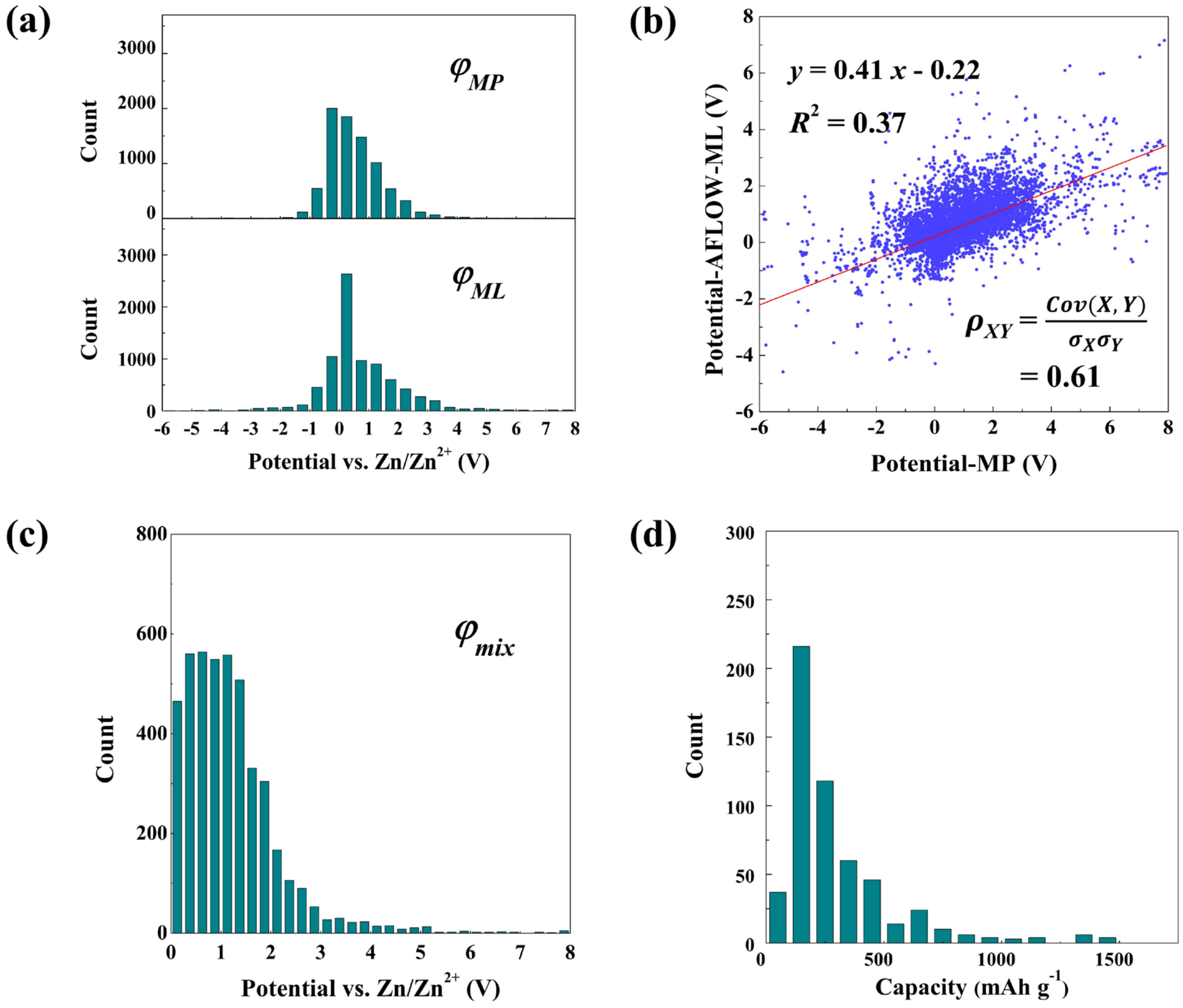

**Fig. 3|** Statistics of the calculated electrochemical potentials for the electrode pairs. (a) Statistical distributions of the potential calculated from MP database (top) and the potential predicted from ML approach (bottom). (b) Parity plot for the potential predicted by ML and calculated from MP, showing a positive correlation. (c) Statistical distribution of the mixing potential $\varphi_{mix}$. (d) Statistical distribution of the capacity, we choose only the polymorph with the lowest energy as the ground state.

The theoretical specific capacity is calculated, using the molecular weight of the cathode material $X$ and the charge transfer in electrochemical reaction Eq.1. It can be observed in Fig.3

(d) that among all the pairs with a predicted potential greater than 0.5 V, ~400 pairs give rise to a theoretical specific capacity >150 mAh/g. Among them, some candidates share the same elements and atomic ratio with different total atom numbers (e.g., for Mn peroxide, they may appear in the form of $MnO_2$, $Mn_2O_4$, and $Mn_3O_6$, etc.). In this study, we choose only the lowest total atom numbers and the polymorph with the lowest energy as the ground state.

After all the screening and filtering, some representative predicted cathode materials are listed in **Fig.4**. It can be observed that some of the experimentally discovered inorganic cathode materials are captured, including $VO_2$[12], $MnO_2$[9], $VS_2$[40], $V_3O_7$[36], $V_2O_3$[37], $V_4O_9$[38], $I_2$[42], and $O_2$[43] etc. Most of the experimentally measured values for these materials agree remarkably well with the predicted potentials by mixing prediction. While $\varphi_{MP}$ and $\varphi_{ML}$ differ from each other as expected and most of the experimental values fall in between the two potentials, which can serve as the uncertainty for the theoretical predictions. Meanwhile, some new promising cathode materials are predicted to give a potential value between 0.5 V to 1.5 V while also give a reasonably high capacity, including $SnPO_4$, $MnPO_4$, $AgPS_2$, $CoP_2O_7$, $MnP_2O_7$, $BaMnF_7$, $BaNi_4O_8$, and $Li_3Fe_2(PO_4)_3$, etc. A full list of all the 66 finalists and their Materials Project identity number (MP-id) for $Zn_aX_b$ and $X$ are given in **Table S2**. It can be seen that the high voltage candidates are mainly in four categories, including oxides, fluorides, sulfides, and phosphates, which shows some similarity with the current high voltage cathode materials for Li-batteries.

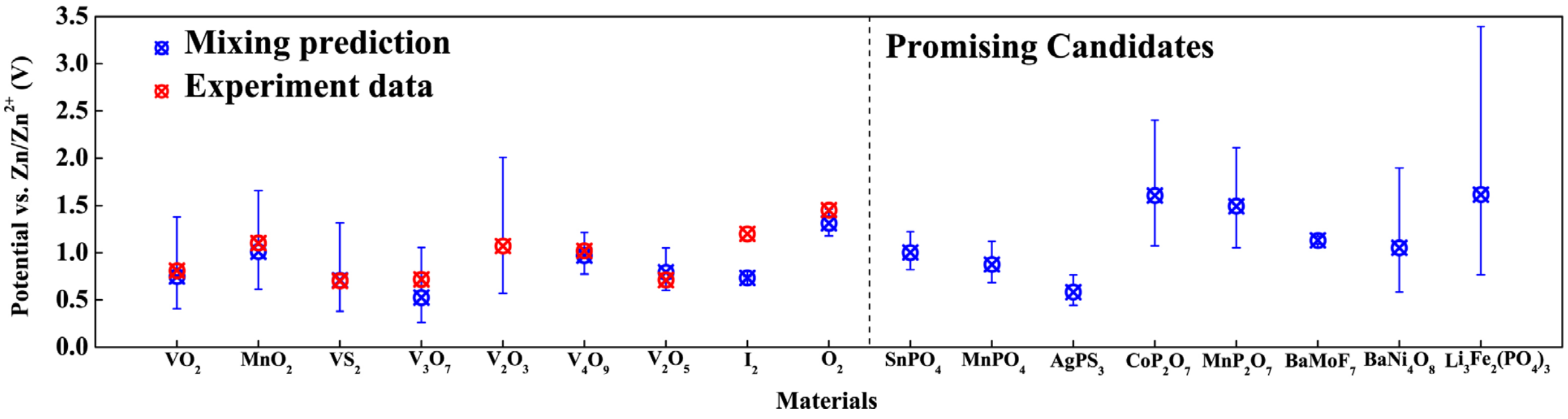

**Fig. 4|** The theoretical prediction with uncertainty for representative candidates from the list of finalists. The results on the left shows the mixing prediction is in consistence with the experiment data for the experimentally discovered inorganic cathodes. Meanwhile, some predicted cathode materials are listed on the right.

In conclusion, we have developed a CGCNN-based ML tool to predict high voltage cathode materials for Zn-ion batteries, which mix the potential data calculated from the MP database and predicted from ML-learned potential by AFLOW database. We screened over ~130,000 inorganic materials and extracted hundreds of candidates that are predicted to have high voltage (over 0.5 V vs $Zn/Zn^{2+}$) and high capacity (over 100 mAh/g). The mixing predictions agree remarkably well with experimental data, while the two potentials can provide an estimate for the uncertainty of the predictions. We further provide possible candidates for high potential Zn battery cathode materials for future experimental exploration. This approach can be widely extended to other battery systems including Li-, Na-, K-, Al- and Mg- based batteries. We hope this study could spur further interest in the applications of ML-based approach for battery materials discovery.

**Notes**

The authors declare no competing financial interest.

**Supporting Information Available**

The Supporting Information is available free of charge on the ACS Publications website at

DOI: XXXX.

Descriptions: **Model and hyperparameters, Predicted electrode pairs**

## ACKNOWLEDGMENTS

Z. H. is supported by a start-up grant from Zhejiang University. This work is supported by the Fundamental Research Funds for the Central Universities.

**TOC graphic**

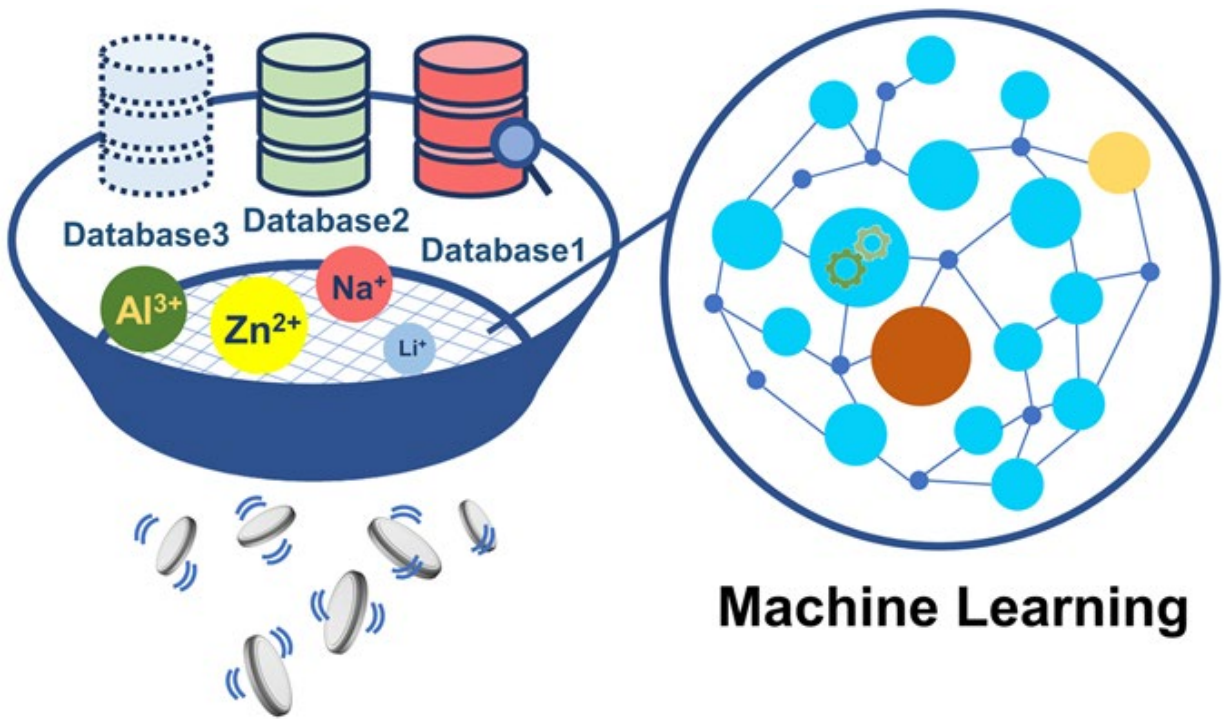

# Supplementary Information: Machine Learning Enabled Prediction of Cathode Materials for Zn ion Battery

Linming Zhou[1,2], Archie Mingze Yao[3], Yongjun Wu[1, 4#], Ziyi Hu[1],

Yuhui Huang[1, #], Zijian Hong[1,2#]

*[1] Lab of dielectric Materials, School of Materials Science and Engineering, Zhejiang University, Hangzhou, Zhejiang 310027, China*

*[2] Hangzhou Global Scientific and Technological Innovation Center, Zhejiang University, Hangzhou, Zhejiang 311200, China*

*[3] Department of Mechanical Engineering, Carnegie Mellon University, Pittsburgh, PA, 15213, USA*

*[4] State Key Laboratory of Silicon Materials, Cyrus Tang Center for Sensor Materials and Applications, School of Materials Science and Engineering, Zhejiang University, Hangzhou 310027, China*

## Model and hyperparameters

In this work, we employ the original crystal graph convolution neural networks (CGCNN) developed by Xie *et al.* [1]. In CGCNN method, the crystal graph consists of two components, atoms features (Group number, Period number, Electronegativity, Covalent radius, Valence electrons, First ionization energy, Electron affinity, Block, Atomic volume) and bond feature (Distance). The crystal graph is connected to the target output through convolution layers: $L_1$ fully connected with the hidden and pooling layers, while $L_2$ fully connected with hidden layers. Using backpropagation and stochastic gradient descent (SGD), we can solve the optimization problem by iteratively update the weights with DFT calculated data from AFLOW database. The main hyperparameters for this study are listed in **Table S1**.

**Table 1**| Main hyperparameters used in CGCNN-DFT-solver training

| Hyperparameters | Value |
|---|---|
| mini-batch size | 256 |
| Initial learning rate | 0.01 |
| momentum | 0.9 |
| weight-decay | 0 |
| optimize method | SGD |
| number of hidden atom features in conv layers | 64 |
| number of hidden features after pooling | 128 |
| number of conv layers | 3 |
| number of hidden layers after pooling | 1 |
| maximum number of neighbors | 12 |
| maximum radius | 8.0 |
| gaussian distance step | 0.2 |
| dropout | 0 |

## Predicted electrode pairs

In this work, we obtain a total of 19197 electrode pairs from the MP database and further screened 8909 electrode pairs without toxic and radioactive elements. The MP-id for the electrode pairs are listed in two files, **total-19197.xlsx** and **normal-8909.xlsx**, which are provided as separate supplementary files. After the final filtering as shown in the main text, we summarize the finalist of predicted 68 cathode pairs, which is given in **Table S2**.

**Table 2**| Finalist of predicted 68 cathode pairs

| Materials Category | $X$ | MP-id | $Zn_aX_b$ | MP-id | $\varphi_{MP}$ (V) | $\varphi_{ML}$ (V) | $\varphi_{Mix}$ (V) | $\varphi_{Exp}$ (V) | Capacity mAh/g |
|---|---|---|---|---|---|---|---|---|---|
| Oxides | $VO_2$ | mvc-13110 | $ZnV_4O_8$ | mvc-13217 | 1.378 | 0.406 | 0.748 | 0.81[2] | 161.496 |
| | $V_3O_7$ | mp-622640 | $ZnV_3O_7$ | mvc-9164 | 1.056 | 0.262 | 0.526 | 0.72[3] | 202.319 |
| | $V_2O_3$ | mvc-846 | $ZnV_4O_6$ | mvc-822 | 2.008 | 0.572 | 1.071 | 1.07[4] | 178.735 |
| | $MnO_2$ | mp-1279979 | $ZnMn_4O_8$ | mvc-172 | 0.614 | 1.658 | 1.009 | 1.10[5] | 154.073 |
| | $CoO_2$ | mp-32686 | $ZnCo_2O_4$ | mp-753489 | 1.745 | 2.140 | 1.932 | / | 294.608 |
| | $V_4O_9$ | mvc-10961 | $ZnV_4O_9$ | mp-1244754 | 1.214 | 0.774 | 0.969 | 1.02[6] | 154.066 |
| | $MoO_2$ | mp-25427 | $ZnMo_2O_4$ | mvc-154 | 0.334 | 1.638 | 0.739 | / | 104.695 |
| | $MoO_3$ | mvc-13534 | $ZnMo_2O_6$ | mvc-2162 | 4.024 | 1.597 | 2.535 | / | 186.116 |
| | $Sn_3O_8$ | mp-1219002 | $Zn_2Sn_3O_8$ | mvc-7701 | 3.055 | 1.050 | 1.791 | / | 221.340 |
| | $Ni_2O_3$ | mvc-894 | $ZnNi_4O_6$ | mvc-876 | 2.481 | 1.224 | 1.743 | / | 161.982 |
| | $Co_9O_{13}$ | mvc-2263 | $Zn_2Co_9O_{13}$ | mvc-2284 | 1.489 | 1.087 | 1.272 | / | 145.122 |
| | $Co_2O_3$ | mvc-852 | $ZnCo_4O_6$ | mvc-823 | 1.654 | 0.892 | 1.215 | / | 161.513 |
| | $SrO_2$ | mp-2697 | $ZnSrO_2$ | mp-5637 | 1.341 | 0.472 | 0.795 | / | 447.909 |
| | $V_2O_5$ | mp-776152 | $ZnV_4O_{10}$ | mp-1314667 | 1.050 | 0.602 | 0.795 | 0.71[7] | 147.290 |
| Fluorides | $SbF_5$ | mvc-3453 | $ZnSbF_5$ | mvc-3462 | 1.337 | 2.521 | 1.836 | / | 247.188 |
| | $VF_4$ | mp-863400 | $ZnVF_4$ | mvc-3525 | 0.762 | 0.897 | 0.827 | / | 422.093 |
| | $CoF_4$ | mvc-13941 | $ZnCoF_4$ | mvc-3380 | 2.805 | 1.264 | 1.883 | / | 397.095 |
| | $MnF_4$ | mp-753574 | $ZnMnF_4$ | mvc-3477 | 2.293 | 1.799 | 2.031 | / | 409.211 |
| | $SnF_4$ | mp-2706 | $ZnSnF_4$ | mvc-3242 | 0.921 | 1.765 | 1.275 | / | 275.178 |
| | $MoF_6$ | mp-557259 | $ZnMoF_6$ | mp-1540855 | 1.154 | 1.707 | 1.404 | / | 255.221 |
| Sulfides | $VS_2$ | mp-9561 | $ZnV_2S_4$ | mvc-14772 | 1.317 | 0.381 | 0.708 | 0.70[8] | 322.993 |
| | $CoS_2$ | mvc-11233 | $ZnCo_4S_8$ | mvc-11285 | 0.620 | 1.156 | 0.846 | / | 108.843 |
| | $MnS_2$ | mp-1018804 | $ZnMn_4S_8$ | mvc-12951 | 0.816 | 4.357 | 1.885 | / | 112.495 |
| | $TiS_2$ | mp-1072192 | $ZnTi_4S_8$ | mvc-11313 | 4.635 | 6.259 | 5.386 | 0.40[9] | 119.597 |
| Phosphates | $MnPO_4$ | mp-540118 | $ZnMn_2(PO_4)_2$ | mvc-10044 | 1.121 | 0.684 | 0.876 | / | 178.703 |
| | $SnPO_4$ | mp-673078 | $ZnSn_2(PO_4)_2$ | mvc-9630 | 1.222 | 0.823 | 1.003 | / | 125.369 |
| | $Fe_9(PO_4)_8$ | mp-704206 | $Zn_3Fe_9(PO_4)_8$ | mp-1216151 | 1.580 | 0.441 | 0.835 | / | 127.328 |
| | $SrV(PO_4)_2$ | mp-1575885 | $ZnSrV\ (PO_4)_2$ | mp-1660296 | 0.938 | 0.843 | 0.889 | / | 163.098 |
| | $SrMn(PO_4)_2$ | mvc-3015 | $ZnSrMn(PO_4)_2$ | mp-1660336 | 2.637 | 1.352 | 1.888 | / | 161.138 |
| | $Li_3Fe_2(PO_4)_3$ | mp-585226 | $ZnLi_3Fe_2(PO_4)_3$ | mp-1223044 | 3.394 | 0.768 | 1.615 | / | 128.354 |
| | $MoTi(PO_4)_3$ | mp-1239324 | $ZnTiMo(PO_4)_3$ | mp-1244626 | 0.660 | 0.357 | 0.485 | / | 124.973 |
| | $TiV(PO_4)_3$ | mvc-9175 | $ZnTiV(PO_4)_3$ | mvc-9187 | 0.523 | 0.308 | 0.401 | / | 139.628 |
| Others | $I_2$ | mp-23153 | $ZnI_2$ | mp-570964 | 0.717 | 0.749 | 0.733 | 1.20[10] | 211.097 |
| | $O_2$ | mp-1180036 | ZnO | mp-2133 | 1.462 | 1.176 | 1.312 | 1.45[11] | 3348.834 |
| | S | mp-7 | $ZnS_2$ | mp-1102743 | 0.616 | 0.813 | 0.708 | / | 835.459 |
| | $MnSe_2$ | mp-21321 | $ZnMn_2Se_4$ | mp-1103546 | 1.146 | 0.709 | 0.902 | / | 125.854 |
| | $Ba(SeO_3)_2$ | mp-1182383 | $ZnBa(SeO_3)_2$ | mp-1205399 | 4.135 | 1.559 | 2.539 | / | 136.944 |
| | $Mn_2NiO_6$ | mp-25285 | $Zn_3Mn_4(NiO_6)_2$ | mp-1222033 | 2.039 | 1.754 | 1.891 | / | 303.773 |
| | $Ni(PO_3)_4$ | mp-504269 | $ZnNi(PO_3)_4$ | mp-540948 | 3.164 | 1.030 | 1.805 | / | 143.036 |
| | $Ni_3(P_2O_7)_2$ | mp-851093 | $ZnNi_3(P_2O_7)_2$ | mvc-881 | 2.100 | 1.320 | 1.665 | / | 102.256 |
| | $BaV_2O_7$ | mp-1201280 | $ZnBaV_2O_7$ | mp-560496 | 3.084 | 0.866 | 1.635 | / | 152.555 |
| | $CoP_2O_7$ | mp-31552 | $ZnCoP_2O_7$ | mp-1567231 | 2.404 | 1.072 | 1.606 | / | 230.073 |
| | $BaGeO_4$ | mp-1190729 | $ZnBaGeO_4$ | mp-757722 | 2.672 | 0.878 | 1.532 | / | 195.567 |
| | $MnP_2O_7$ | mp-770542 | $ZnMnP_2O_7$ | mvc-1238 | 2.113 | 1.053 | 1.491 | / | 234.089 |
| | $Fe_3(P_2O_7)_2$ | mp-1101565 | $ZnFe_3(P_2O_7)_2$ | mvc-901 | 1.107 | 1.602 | 1.332 | / | 103.951 |
| | $Mn(GeO_3)_2$ | mvc-9216 | $ZnMn(GeO_3)_2$ | mp-1567247 | 2.496 | 0.685 | 1.307 | / | 180.877 |
| | $NiP_2O_7$ | mp-689834 | $ZnNiP_2O_7$ | mp-1639312 | 2.810 | 0.538 | 1.230 | / | 230.310 |
| | $Mn_3(P_2O_7)_2$ | mp-772322 | $ZnMn_3(P_2O_7)_2$ | mvc-870 | 2.346 | 0.644 | 1.230 | / | 104.502 |
| | $Co_3(P_2O_7)_2$ | mp-1638872 | $ZnCo_3(P_2O_7)_2$ | mvc-884 | 1.990 | 0.646 | 1.134 | / | 102.115 |
| | $LiV_3O_8$ | mp-27503 | $Zn_2LiV_3O_8$ | mp-1280709 | 1.089 | 1.076 | 1.083 | / | 372.381 |
| | $Ba(NiO_2)_4$ | mp-18991 | $ZnBa(NiO_2)_4$ | mvc-945 | 0.584 | 1.897 | 1.052 | / | 107.137 |
| | $Fe_2(MoO_4)_3$ | mvc-8023 | $Zn_3Fe_4(MoO_4)_6$ | mvc-8032 | 1.566 | 0.698 | 1.045 | / | 135.870 |
| | $Mn_3Nb_2O_9$ | mvc-12227 | $ZnMn_3Nb_2O_9$ | mp-557349 | 1.448 | 0.684 | 0.995 | / | 108.322 |
| | $CuMo_2O_7$ | mp-1244839 | $ZnCuMo_2O_7$ | mvc-6675 | 0.813 | 1.030 | 0.915 | / | 145.823 |
| | $V_3P_3O_{13}$ | mvc-10619 | $ZnV_3P_3O_{13}$ | mvc-10544 | 1.748 | 0.475 | 0.911 | / | 118.082 |
| | $MoSi_2O_7$ | mp-1143109 | $Zn_3Si_4(MoO_7)_2$ | mp-1265617 | 1.386 | 0.569 | 0.888 | / | 304.299 |
| | $VPO_5$ | mp-25265 | $ZnV_2(PO_5)_2$ | mvc-8904 | 0.841 | 0.938 | 0.888 | / | 165.455 |
| | $CuMoO_4$ | mp-619545 | $ZnCu_2(MoO_4)_2$ | mvc-10467 | 0.745 | 0.843 | 0.792 | / | 119.871 |
| | $V_2P_2O_9$ | mp-628951 | $ZnV_2P_2O_9$ | mvc-10173 | 0.967 | 0.601 | 0.762 | / | 174.055 |
| | $SnP_2O_7$ | mvc-1016 | $ZnSnP_2O_7$ | mvc-5610 | 1.236 | 0.440 | 0.738 | / | 183.078 |
| | $Si_4SnO_{10}$ | mp-1143317 | $ZnSi_4SnO_{10}$ | mvc-8492 | 0.995 | 0.460 | 0.676 | / | 137.012 |
| | $Mo_2P_2O_9$ | mvc-10108 | $ZnMo_2P_2O_9$ | mvc-10165 | 1.203 | 0.375 | 0.672 | / | 134.679 |
| | $AgPS_3$ | mp-5470 | $ZnAg_2(PS_3)_2$ | mp-1207470 | 0.766 | 0.444 | 0.583 | / | 113.978 |
| | $VP_2O_7$ | mp-26818 | $ZnVP_2O_7$ | mp-1639659 | 0.501 | 0.522 | 0.511 | / | 238.248 |
| | $VNiP_2O_9$ | mvc-12312 | $ZnVNiP_2O_9$ | mvc-8267 | 1.998 | 0.445 | 0.943 | / | 169.780 |
| | $LaFeSbO_6$ | mvc-9014 | $ZnLaFeSbO_6$ | mvc-9031 | 0.668 | 0.338 | 0.475 | / | 129.884 |
| | $LaFeMoO_6$ | mvc-8986 | $ZnLaFeMoO_6$ | mp-1641568 | 1.643 | 0.113 | 0.431 | / | 138.557 |
| | $NiMoP_2O_9$ | mvc-8474 | $ZnNiMoP_2O_9$ | mvc-8300 | 1.064 | 0.158 | 0.410 | / | 148.592 |